\documentclass[useAMS,graphicx,usegraphicx,usenatbib,11pt]{mn2e}
\usepackage{subfigure}
\usepackage[fleqn]{amsmath}
\usepackage{amssymb}
\usepackage[english]{babel}  
\usepackage{array}
\usepackage{graphicx}
\usepackage{multicol}
\usepackage{times}
\usepackage{bm}

\newcommand{\rhrj}{\mathcal{R}}

\newcommand{\rh}{r_{\rm h}}

\newcommand{\rhl}{r_{\rm hl}}

\newcommand{\rj}{r_{\rm J}}

\newcommand{\dr}{{\rm d}}
\newcommand{\msun}{{\rm M}_\odot}
\newcommand{\rg}{R_{\rm G}}
\newcommand{\rgmin}{R_{\rm G}^{\rm min}}
\newcommand{\rgmax}{R_{\rm G}^{\rm max}}
\newcommand{\vg}{V_{\rm G}}
\newcommand{\mi}{M_0}
\newcommand{\mdot}{\dot{M}}
\newcommand{\rhmax}{r_{\rm h}^{\rm max}}
\newcommand{\trh}{\tau_{\rm rh}}
\newcommand{\myr}{\mbox{Myr}}
\newcommand{\mlo}{M_{\rm low}}
\newcommand{\mup}{M_{\rm up}}
\newcommand{\rpeak}{\hat{r}_{\rm h}}

\newcommand{\mpeak}{\hat{M}}
\newcommand{\tdis}{t_{\rm dis}}
\newcommand{\kpc}{\mbox{kpc}} 
\newcommand{\kms}{\mbox{km\,s}^{-1}}

\title[The initial conditions of GCs from their radii]{Constraining the initial conditions of globular clusters using their radius distribution}
\author[P.~E.~R.~Alexander \& M. Gieles]{Poul~E.~R.~Alexander$^{1}$\thanks{e-mail: pera@ast.cam.ac.uk} and Mark~Gieles$^{1,2}$
\\ $^1$Institute of Astronomy, University of Cambridge, Madingley Road, Cambridge, CB3 0HA, UK
\\ $^2$Department of Physics, University of Surrey, Guildford, GU2 7XH, UK}
\date{Accepted 2013 February 09; Received 2013 February 08; in original form: 2013 January 21}

\begin{document} 

\maketitle

\begin{abstract}
Studies of extra-galactic globular clusters have shown that the peak size of the globular cluster (GC) radius distribution (RD) depends only weakly on galactic environment, and can be used as a standard ruler. We model RDs of GC populations using a simple prescription for a Hubble time of relaxation driven evolution of cluster mass and radius, and explore the conditions under which the RD can be used as a standard ruler. We consider a power-law cluster initial mass function (CIMF) with and without an exponential truncation, and focus in particular on a flat and a  steep CIMF (power-law indices of $0$ and $-2$, respectively). For the initial half-mass radii at birth we adopt either Roche-lobe filling conditions  (`filling', meaning that the ratio of half-mass to Jacobi radius is approximately $\rh/\rj \simeq 0.15$) or strongly Roche-lobe under-filling  conditions (`under-filling', implying that initially $\rh/\rj \ll 0.15$). Assuming a constant orbital velocity about the galaxy centre we find for a steep CIMF that the typical half-light radius scales with galactocentric radius $\rg$ as $\rg^{1/3}$. This weak scaling is consistent with observations, but this scenario has the (well known) problem that too many low-mass clusters survive. A flat CIMF with `filling' initial conditions results in the correct mass function at old ages, but with too many large (massive) clusters at large $\rg$. An `under-filling' GC population with a flat CIMF also results in the correct mass function, and can also successfully reproduce the shape of the RD, with a peak size that is (almost) independent of $\rg$. In this case, the peak size depends (almost) only on the peak mass of the GC mass function. The (near) universality of the GC RD is therefore because of the (near) universality of the CIMF. There are some extended GCs in the outer halo of the Milky Way that cannot be explained by this model.
\end{abstract}
\begin{keywords}
galaxies: star clusters -- globular clusters: general
\end{keywords}

\section{Introduction} 
It has long been known that the shape of globular cluster (GC) luminosity function is relatively insensitive to galactic environment. Accordingly, the peak luminosity of $M_V \simeq -7$ is often used as a standard candle \citep{SS1927, 2012Ap&SS.341..195R}. It is however not known whether this typical luminosity is the result of dynamical evolution \citep{FZ2001} or the outcome  of the cluster formation process \citep*[see][for reviews]{2006ARA&A..44..193B, 2010ARA&A..48..431P}.

\citet{JCBea2005} showed that the  half-light radii ($\rhl$) of GCs are also sufficiently independent of galactic environment to be used as a standard ruler. Typical radii of $\rhl\simeq3\,$pc are found for clusters in the Milky Way \citep[][2010 version]{H1996} and other galaxies \citep{KW2001,JCBea2005,PFG2011}. Within galaxies the typical radii depend only weakly on galactocentric radius $\rg$. For the Milky Way GCs \citet{vdB1991} and \citet{2000ApJ...539..618M} find $\rg^{0.5}$ and $\rg^{0.4}$, respectively. Similar and shallower correlations are found for clusters in external galaxies:  $\rh \propto \rg^{0.38}$ (in the Sombrero galaxy; \citealt{SLSB2006}), $\rh\propto\rg^{0.17}$ (for the inner regions of NGC 5128; \citealt{GW2007}), $\rh \propto R_{\rm G}^{0.11}$ (in a sample of six ellipticals; \citealt{H2009}). If  clusters fill their Jacobi radius $\rj$, their radii scale with their mass $M$ and their orbital frequency $\Omega$ as $M^{1/3}\Omega^{-2/3}$. Hence, for a constant $M$ and  orbital velocity $\vg$ we expect $\rhl\propto\rg^{2/3}$, i.e. a stronger dependence on $\rg$ than  observed.

The objective of this work is to understand how the shape of the cluster radius distribution (RD) depends on the cluster initial mass function (CIMF), the distribution of clusters within the galaxy and the initial cluster sizes. This will allow us to use the RD to put additional constraints on the initial conditions of GCs.

The structure of this paper is as follows: In Section~\ref{s:1rg} we discuss the RD arising for a synthetic sample of GCs at one $\rg$. In Section~\ref{s:rgd} we consider an entire population of synthetic clusters and compare our results with the Milky Way GC system. We finally compare our results to previous studies and provide a summary in Section~\ref{s:conc}.

\section{The radius distribution of clusters in a single tidal environment}
\label{s:1rg}
We start by considering the radius distribution (RD), i.e. the number clusters with a half-mass radius between $\rh$ and $\rh+\dr\rh$,  of clusters evolving in a fixed tidal environment (for example, evolving at a constant $\rg$ within a galaxy halo).  For this situation the RD can be related to the cluster initial mass function (CIMF, i.e. the number of clusters with initial mass in the interval $\mi$ and $\mi+\dr\mi$) as

\begin{align}
\frac{\dr N}{\dr \rh}(\rg) = 
\frac{\dr N}{\dr \mi}\left|\frac{\partial \mi}{\partial M}\right|\left|\frac{\partial M}{\partial \rh}(\rg)\right|.
\label{eq:con-num}
\end{align}

Here $|\partial \mi/\partial M|$  and $|\partial M/\partial \rh(\rg)|$ depend on the mass and radius evolution, respectively. These will be discussed in the following sections.
\subsection{Initial mass function and mass evolution}
\label{ssec:mt}
In this section we assume that the CIMF is a power-law distribution: $\dr N/\dr\mi \propto \mi^\alpha$ in the range $\mlo\le\mi\le\mup$. In Section~\ref{s:rgd} we additionally consider  \citet{S1976} distributions. We aim to reproduce the GC population of a Milky Way like galaxy, and therefore adopt a constant $\vg=220\,\kms$ such that $\Omega \propto \rg^{-1}$. Throughout this work we assume that clusters evolve with a constant mass-loss rate that depends on the galactocentric radius as $\mdot\rg = -20\,\msun\myr^{-1}\kpc$ \citep*{GHZ2011}. The mass then evolves as 
\begin{equation}
M = \mi - \Delta, \mbox{with } \Delta \equiv -\mdot t\propto \rg^{-1}.
\end{equation}
Here, $t$ is the age of the cluster, for which we adopt 13 Gyr throughout this work. We then have $|{\partial \mi}/{\partial M }|= 1$.  The total lifetime, or dissolution time, is $\tdis =-\mi/\mdot$.   For the final term of equation~(\ref{eq:con-num}), we need to know how the radii of clusters depend on $M$, for which we compare two different scenarios. In Section~\ref{ssec:filling}, we consider Roche-lobe filling clusters, whereby the ratio of half-mass to Jacobi radius ($\rhrj \equiv \rh/\rj$) is constant and approximately the \citet{H1961} value, $\rhrj  = 0.145$. In Section~\ref{ssec:ufilling}, we consider initially Roche-lobe under-filling clusters, where initially $\rhrj = 0$ and $\rhrj$ increases during the evolution.

\subsection{Roche-lobe filling clusters}
\label{ssec:filling}
We first consider the scenario whereby all clusters fill their Roche-lobe from birth.  Such clusters will evolve at a constant density whilst losing mass \citep{H1961} and so

\begin{equation}
M = A\rh^3, \mbox{ with }A\equiv\frac{2\Omega^2}{G\rhrj^3}\propto\rg^{-2},
\label{eq:rhf}
\end{equation}

where $G$ is the gravitational constant.  From this we find \textbf{$|\partial M/\partial\rh(\rg)| \propto\rh^2\rg^{-2}$}, which we combine with equation~(\ref{eq:con-num}) to derive the RD
\begin{align}
\frac{\dr N}{\dr \rh}(\rg) &\propto \left(A\rh^3 + \Delta\right)^\alpha\rh^2\rg^{-2}\notag \\
&\propto
\begin{cases}
 \rh^2\rg^{-(2+\alpha)}          &,0 \le A\rh^3 \lesssim \Delta,\\
 \rh^{3\alpha+2}\rg^{-2(\alpha+1)}  &,\ \Delta\lesssim A\rh^3\le\mup.
\end{cases}
\label{eq:contract}
\end{align}
For $\alpha=0$ the RD is a single power-law with slope $+2$, while for $\alpha\ne0$ it is a double power-law distribution: at small radii the shape of the RD is independent of the shape of the CIMF and only depends on how clusters lose mass. This is because the Globular Cluster Mass Function (GCMF) evolves to $\dr N/\dr M\simeq\, $constant at low masses, regardless of the shape of the CIMF \citep{H1961, FZ2001}. At large radii the shape of the RD depends on the index of the CIMF, for which we find two regimes: firstly, for $\alpha <-2/3$ the RD is an decreasing function of $\rh$ and the peak of the distribution occurs where the two power-laws join, i.e. at $\rpeak \simeq (\Delta/A )^{1/3} \propto \rg^{1/3}$, where we introduce the symbol $\rpeak$ to denote the mode of the RD (most probable radius). For $\alpha>-2/3$ the RD rises until the largest radius present, which is the radius of the most massive cluster ($\mup$) and thus $\rpeak \propto \rg^{2/3}$.


\subsection{Roche-lobe under-filling clusters}
\label{ssec:ufilling}
We next consider clusters that are born Roche-lobe under-filling. We use the model of \citet{GHZ2011} for the relaxation driven evolution of $\rh$ as a function of $\mi$ and $\rg$, in which clusters are assumed to form with a half-mass density much greater than the tidal density. In the first (roughly)  half of their life clusters expand and become increasingly Roche-lobe filling, while in the second half they contract at a constant density, and near final dissolution the radius evolves as in equation~(\ref{eq:rhf}) from Section~\ref{ssec:filling}.

\begin{figure}
\centering
\includegraphics[width=85mm]{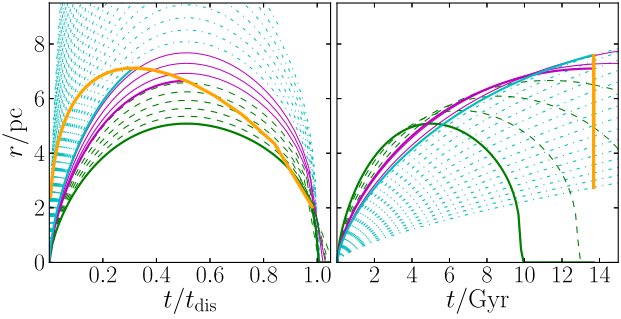}
\caption{Radius evolution of clusters of different masses
  ($10^4\,\msun < M < 10^7\,\msun$) at $\rg = 8.5\,\kpc$, $\vg =
  220\,\kms$, as a function of fractional dissolution time (left
  panel) and physical time (right panel). The thick/orange line represents
  the $\rh$ of clusters after a Hubble time, and is curved in the left panel
  on account of the longer $\tdis$ of more massive clusters. The green/dashed,
  magenta/solid and cyan/dot-dash lines represent the evolutionary tracks of individual
  (sample) clusters in states (i), (ii) and (iii), respectively. These tracks
  are thicker for evolution that has occurred during a Hubble time, that is, on
the left of the orange line.}
\label{f:single_rg_evo}
\end{figure}

The functional form for the evolution of $\rh$ is (equation~(26) in \citealt{GHZ2011})
\begin{align}
  \rh&= 
  \left(\frac{M}{A}\right)^{1/3}\left(1-\left[\frac{M}{\mi}\right]^{5/2}\right)^{2/3}.
 \label{eq:rm}
\end{align}
Note that $A=A(\rg)$, $M = M(\mi,\Delta)$ and $\Delta=\Delta(\rg, t)$. In the late stages of evolution ($M\ll\mi$), this function converges to the Roche-lobe filling relation (equation~\ref{eq:rhf}). Meanwhile, for $M\simeq \mi$ we find $\rh\propto\mi^{-1/3}t^{2/3}$, which follows from the fact that in the expansion phase clusters evolve toward a constant ratio of $\trh/t$ with $\trh\propto(M\rh^3)^{1/2}$ \citep[assuming a constant mean stellar mass and Coulomb logarithm,][]{1971ApJ...164..399S}. If there are clusters in both the expansion and contraction phases, the size of the largest cluster ($\rhmax$) is not that of the most massive cluster, as was the case for Roche-lobe filling clusters (Section~\ref{ssec:filling}). To find $\rhmax$ in a population of clusters with different $\mi$ at a given $\rg$ and $t$, we numerically solve $\dr \rh/\dr \mi=0$ (equation~\ref{eq:rm}), and find $\mi(\rh=\rhmax) \simeq 3.175\Delta$.  We note that this is \emph{not} the mass of the cluster that has just reached its maximum size, but instead a slightly more massive cluster that is still expanding (see Fig.~\ref{f:single_rg_evo}). This is because the maximum $\rh$ is reached roughly half-way the total lifetime \citep{GHZ2011}, so the cluster that has reached its maximum size has an initial mass of $\mi\simeq2\Delta$. From Fig.~\ref{f:single_rg_evo} we also see that clusters can be placed in three categories, which we base on the their $\mi$:

\begin{enumerate}
\item{ $\mi \le 2\Delta$ -- clusters have past their individual  maximum $\rh$ and are now approximately Roche-lobe filling (dashed/green  tracks).}
\item{$2\Delta<\mi\le3.175\Delta$ -- clusters that are expanding toward their individual maximum $\rh$, and show a positive correlation between mass and $\rh$ (solid/magenta tracks).}
\item{$\mi>3.175\Delta$ -- clusters that are expanding, and have a negative correlation between $M$ and $\rh$ (dot-dashed/cyan tracks).}
\end{enumerate}

We cannot derive an easy analytical expression for the RD, because we can not invert the relation $\rh(\mi)$ (equation~\ref{eq:rm}).  We are however able to give the behaviour of the RD at the extremes of $\rh$.  For expanding clusters, we use $\rh\propto\mi^{-1/3}$ and $M\simeq\mi$ to find $\left|\partial M/\dr \rh\right| \propto \rh^{-4}$, while contracting clusters approximately follow the same relationships as in Section~\ref{ssec:filling}. We substitute these expressions into equation~(\ref{eq:con-num}) and find
\begin{align}
\frac{\dr N}{\dr \rh}(\rg) &\propto 
\begin{cases}
 \rh^2\rg^{-(2+\alpha)}              &, 0\le\rh\lesssim\rhmax,\\
\rh^{-(3\alpha +4)}
&, \rh(\mup)\le\rh\lesssim\rhmax.
\end{cases}
\label{eq:expand}
\end{align}
Here we assumed that clusters with $M=\mlo$ have dissolved already and that $\rh$ shrinks to 0 near dissolution. These two parts of the RD both occupy the region $0\lesssim\rh\lesssim\rhmax\propto \rg^{1/3}$.  The first approximation applies to the clusters whose $\rh$ has approximately reached the Roche-lobe filling condition ($\rhrj \simeq 0.145$), i.e. $\mi \lesssim3.175 \Delta$, while the second approximation holds for the expanding clusters ($\mi\gtrsim 3.175\Delta$). For $\alpha=-2$ we find that both extremes of the RD behave as $\rh^2$. For very high $\mup$ (where $\rh(\mup)\simeq0$), this RD has the same shape as the RD for Roche-lobe filling clusters with $\alpha=0$ (Section~\ref{ssec:filling}), but has a smaller $\rhmax$.

\begin{figure}
\centering
\includegraphics[width=80mm,height=100mm]{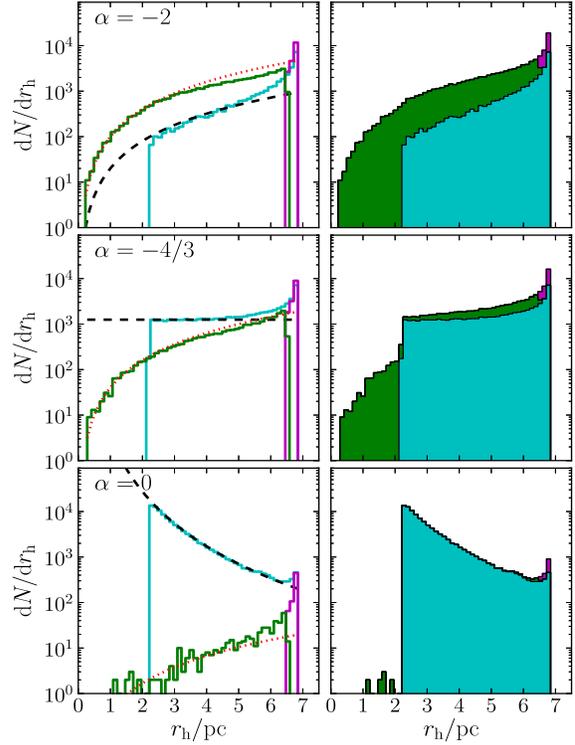}
\caption{Monte Carlo simulations of clusters with different CIMF slopes ($\alpha$) at $\rg= 8.5\,$kpc with $\vg = 220\,\kms$. The colour coding is the same as in Fig.~\ref{f:single_rg_evo} (green = stage (i) clusters, magenta = stage (ii) clusters, cyan = stage (iii) clusters). In the left panels, clusters in the three phases of evolution are shown separately, while they are added in the right panels. The dashed/black line represents the expected slope of expanding clusters while the dotted/red line corresponds to the expected slopes of approximately Roche-lobe filling (contracting) clusters (equation~\ref{eq:expand}). The sample size is $10^5$ clusters and clusters are evolved for 13\,Gyr.}
\label{f:single_rg_pops}
\end{figure}

As for the Roche-lobe filling case, the mode of the RD is either at the radius of the most massive cluster or at the radius of the cluster with $\mi\simeq\rm{few}\times\Delta$, depending on whether the RD of  expanding clusters is rising or falling. The difference with the Roche-lobe filling case is that $\rh(\mup)$ is smaller than $\rhmax$. For $\alpha>-4/3$ the RD of the Roche-lobe filling clusters is rising and $\rpeak=\rhmax\propto \rg^{1/3}$.  For $\alpha<-4/3$ this part of the RD is a declining function, and if the most massive cluster is still expanding the RD peaks at the radius of the most massive cluster: $\rpeak = \rh(\mup) < \rhmax=\,$constant (Fig.~\ref{f:single_rg_pops}). This is the most straightforward way to explain a near universal radius scale for GCs. The condition that the most massive clusters have not yet expanded until their $\rh \simeq \rhrj\rj$ is not satisfied in strong tidal fields, and will be discussed further in Section~\ref{ssec:D}.

The RD is computed for Roche-lobe under-filling clusters and for three values of $\alpha$ using a Monte Carlo approach (Fig.~\ref{f:single_rg_pops}).  The power-law behaviours at the expanding and contracting extremes are shown as dashed/black and dotted/red lines, respectively. The power-laws do not follow the Monte Carlo data where $\rh \simeq \rhmax$, as here $\mi \simeq 3.175\Delta$ and mass loss is neither dominant nor negligible.

Now we have explored the different scenario for clusters at a single
$\rg$ we will consider GC populations in a Milky Way type galaxy in
the next section.

\section{Radius distribution of a cluster system}
\label{s:rgd}
\begin{figure*}
\centering
\includegraphics[width=170mm]{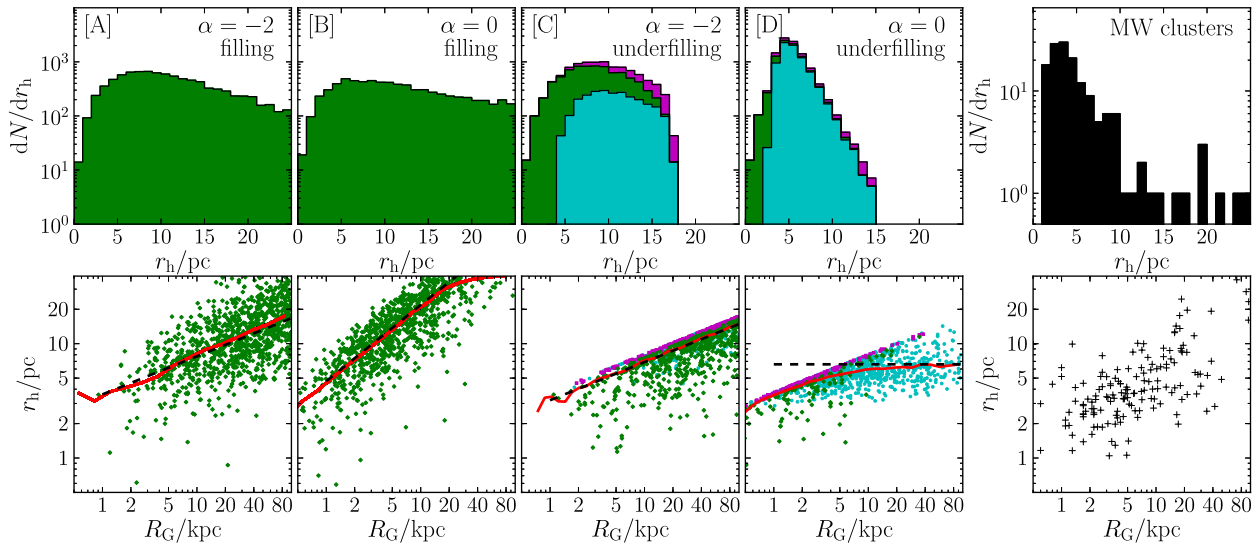}
\caption{Monte Carlo samples of $N = 10^4$ (surviving) star clusters at $t =
  {13}$\,Gyr within an isothermal galactic halo. Clusters are simulated with a
  \citet{H1990} distribution of galactocentric radii and a \citet{S1976}
  CIMF (see text for parameters). In the top row cyan,
  magenta and green regions correspond to clusters undergoing the
  three evolutionary stages outlined in Section~\ref{s:1rg},
  whilst in the bottom row individual sample clusters are denoted by
  green diamonds (stage (i) clusters), magenta squares (stage (ii)
  clusters), or cyan dots (stage (iii) clusters). The left hand panels
  represent scenarios A, B, C, and D as labeled, while the rightmost figure corresponds to the present day Milky Way GC population compiled from \citet[][2010
    version]{H1996}. The red solid line in the bottom row is the median
  $\rh$ as a function of $\rg$, while the black (dotted) lines
  represent $\rh=\, {\rm const}, \rh \propto \rg^{1/3}$, and $\rh \propto \rg^{2/3}$ as appropriate.}
\label{f:rg_pops}
\end{figure*}

In this section we consider the RD of clusters in a Galactic  halo, i.e. with a range of $\rg$.  The RD can be found be multiplying the RD at a single $\rg$ with the number of clusters at each $\rg$ $(\dr N/\dr \rg =4\pi\rg^2 n(\rg)$, where $n(\rg)$ is the number density of clusters in the halo) and integrating over all $\rg$

\begin{align}
\frac{\dr N}{\dr \rh} = 
\int_{\rgmin}^{\rgmax}\frac{\dr N}{\dr \rh}(\rg)\frac{\dr N}{\dr\rg}\dr\rg.
\label{eq:dndrtot}
\end{align}
The analytic evaluation of this integration gives results only for
a limited number of  cases, so we proceed by generating RDs
in a Monte Carlo fashion. We therefore perform a number of simulations, for which we adopt a \citet{H1990} model (i.e., a double power-law density profile with an $r^{-1}$ central cusp and an $r^{-4}$ outer halo) with a scale radius $a = 5$\,kpc and a truncation at $\rg=100\,\kpc$ for the distribution of clusters in the Galaxy ($n(\rg)$), which is a reasonable approximation for GCs in the Milky Way \citep{2000ApJ...539..618M}.
For the CIMF we use a \citet{S1976} function: $\dr N/\dr \mi \propto
\mi^{\alpha}\exp(-\mi/M_*)$, with $M_* = 10^6\,\msun$ and $\mlo =
100\,\msun$. For the slope we consider $\alpha=-2$, i.e. what is found
for young clusters \citep{1999ApJ...527L..81Z, 2010ARA&A..48..431P}
and $\alpha=0$, which describes the MF of old clusters in the Milky
Way and early type galaxies in the Virgo Cluster \citep{JMCF2007}. 
 A CIMF with $\alpha=0$ is invariant under mass-loss \citep{JMCF2007} so at an age of 13~Gyr the GCMF has the same shape as the CIMF.
The resulting RDs and the $\rpeak(\rg)$ relations for these four cases
are shown in Fig.~\ref{f:rg_pops} and are discussed below. We compare
all these simulations to the $\rh$ values of Milky Way GCs compiled
from the \citep[][2010 version]{H1996} catalogue. Values for $\rh$ are derived by multiplying the observed half-light radii by 4/3 to correct for the effect of projection.
 
\subsection{Scenario A: Roche-lobe filling clusters and $\bm{\alpha=-2}$} We first consider Roche-lobe filling clusters and a CIMF with slope $\alpha= -2$. We recall from Section~\ref{ssec:filling} that for Roche-lobe filling clusters there is a critical value for the CIMF slope of $\alpha=-2/3$:  a steeper CIMF gives rise to a double power-law RD, rising at small radii and falling at large radii. The total RD is the sum of many such functions, and therefore has a (logarithmic) slope of +2 at small radii. At large radii the integration of equation~(\ref{eq:dndrtot}) gives a slope of $3\alpha+3.5 = -2.5$. The typical radius scales as $\rpeak\propto \rg^{1/3}$. We recall from Section~\ref{ssec:filling} that this is due to the scaling of the break in the double power-law RD with $\rg$.  This scaling is close to what is found for Milky Way GCs, but in this scenario there are too many large clusters (see panel~A, Fig.~\ref{f:rg_pops}). In addition, too many low-mass clusters survive at large $\rg$ in this scenario \citep{2001MNRAS.322..247V, GB2008}.

\subsection{Scenario B: Roche-lobe filling clusters and $\bm{\alpha=0}$}
For Roche-lobe filling clusters and $\alpha=0$ the RD at a single
$\rg$ is simply a $+2$ power-law.  In this case the integration of
equation~(\ref{eq:dndrtot}) can be performed, with the result that at large
radii we find $\dr N/\dr\rh \propto \rh^{-1}$. Because of this slow decline,
this scenario produces even more large clusters than scenario A.
However, owing to the invariance of this CIMF under the influence of
mass-loss, this scenario does predict the correct GCMF \citep[meaning
  that the peak mass, $\mpeak$, is independent of $\rg$,][]{2000MNRAS.318..841V}. 
 Because $\rpeak$ scales with the Jacobi radius of the most massive cluster (Section~\ref{ssec:filling}) at any given $\rg$, we find $\rpeak \propto \rg^{2/3}$ (see panel~B, Fig.~\ref{f:rg_pops}), which is a stronger correlation than found for Galactic clusters. We note that this is the only scenario our of the four we considered for which the $\rg^{2/3}$ scaling is found.

\subsection{Scenario C: Roche-lobe under-filling and $\bm{\alpha=-2}$}
For under-filling clusters with a steep CIMF ($\alpha =-2$) both the
expanding and contracting clusters have a RD that behaves as a $+2$
power-law (just as in scenario B). However, the total RD is different
  because the typical radius scales with $\rpeak
  \propto \rhmax$ and therefore $\rpeak\propto \rg^{1/3}$ (Section~\ref{ssec:ufilling}). Accordingly, there are fewer large clusters in this scenario, because high mass clusters are still expanding to their tidal boundary and are smaller than clusters of equivalent
mass in scenario B.
\subsection{Scenario D: Roche-lobe under-filling and $\bm{\alpha=0}$}
\label{ssec:D}
In the final scenario (under-filling clusters with $\alpha>-4/3$) the
sample population contains a fraction of expanding clusters, which is larger for
higher values of $\alpha$ (Fig.~\ref{f:single_rg_evo}). The radii of expanding clusters are independent of their local tidal field and also independent of their initial radii \citep{H1965,2010MNRAS.408L..16G}. This is the case for $\rg\gtrsim5\,$kpc (panel D, Fig. 3). The typical $\rh$ is then simply the radius of the cluster with the typical mass. For $\rg < 5\,$kpc, the majority of clusters have expanded to become Roche-lobe filling and are therefore described by scenario B with a typical radius scaling as $\rpeak\propto\rg^{1/3}$. The number of clusters in this regime depends on the slope of the CIMF and the Hernquist scale radius.  This scenario is the only scenario that produces a clear peak in the radius distribution below 5\,pc, and the only scenario to produce a rapid decline in the RD at high $\rh$ consistent with Milky Way GCs. 

\subsection{Comparison to Milky Way GCs}

In the right panels of Fig.~\ref{f:rg_pops} we show the RD and the individual $\rh$ against $\rg$ values of the MW GCs. The width of the Milky Way GC RD is narrow, with radii rarely in excess of 20\,pc (90\% of the GCs have radii $<10\,$pc, \citealt[][2010 version]{H1996}). Comparing this to the models we see that in Scenarios A, B there are too many clusters with $\rh \gtrsim 20\,$pc (see Fig.~\ref{f:rg_pops}), especially in scenario B, where we find clusters with size up to $\rh\sim100$\,pc. Scenarios C and D both show a narrower range of $\rh$, in better agreement with that found for Milky Way GCs, with scenario D providing a very comparable shape overall.

 Several MW GCs at high $\rg$ are larger than those predicted. These outlying clusters cannot be explained by our simple model for cluster evolution, and so may result from effects neglected by this study (e.g., these clusters may be accreted, or perturbed by the tidal fields of external galaxies). The scaling of $\rpeak$ with $\rg$ of the Milky Way GCs shows considerable scatter, but is approximately $\rh \propto \rg^{0.4}$,  similar to the scalings of scenarios A and C and shallower than scenario B. However, the majority of Milky Way clusters are found within the innermost $\sim 10\,$kpc. In this regime the radii in scenario D are similar to those in C. It is therefore difficult to satisfactorily distinguish which scenario optimally represents the $\rh$ scaling of Milky Way clusters, but combined with the superior match of the RD shape we conclude that scenario D is preferred. The $\rh(\rg)$ relationship of scenario D is also similar to that described for M87 in \citet{WSH2012}. We therefore conclude that the good match of the shape of the RD in scenario D suggests that most clusters form Roche-lobe under-filling, and with a relatively flat CIMF, similar to the present day GCMF.

\vspace{-2mm}
\section{\textbf{Summary and discussion}}
\label{s:conc}

We have investigated the radius distribution (RD) of globular cluster (GC) populations in Milky Way-type haloes for a variety of initial conditions for the masses and radii. We find that Roche-lobe filling initial conditions results in too many large clusters ($\gtrsim 20\,$pc) at large $\rg$, for both steep and shallow cluster initial mass function (CIMFs). Models in which clusters are initially Roche-lobe under-filling and form with a steep CIMF show a scaling of $\rpeak\propto\rg^{1/3}$, roughly consistent with Milky Way GCs, but the shape of the synthetic RD is wider than the observed one. The mode of the half-mass RD of a GC population is independent of Galactic environment only if clusters form Roche-lobe under-filling, and with a relatively flat CIMF. 

Alternative models to explain the characteristic GC radius of a few pc exist \citep[e.g.][]{JCBea2005,HSFB2010}. In these models the present day size is the result of the star formation process or the gas expulsion scenario,  respectively. The subsequent Hubble time of evolution is assumed not to significantly change the radii of GCs. In our model scenario D, however, the typical GC size is the natural outcome of a Hubble time of dynamical relaxation and is not sensitive to the exact initial conditions or early evolution since the relaxation driven expansion erases the initial properties of the clusters \citep[in terms of $\rh$,][]{H1965}.

In addition to these works, \citet{MHS2012} show (via $N$-body studies)  that the radius is set ``by clusters that are originally massive enough to survive mass loss due to tidal stripping''. Thus, the present day radii in their model depends only on clusters' ability to survive in the Galactic tidal field.  They consider clusters at different $\rg$, all with the same mass and $\rh$. Their final $\rh$ vs. $\rg$ relation (their Fig.~3) is very similar to our $\rh(\rg)$ relation found in our scenario D. The radii of their clusters at large $\rg$ are larger, because they consider clusters with lower masses which have expanded more \citep{H1965}.

In contrast to our results, \citet{SKYK2013} find from a large suite of Fokker-Planck simulations that models in which GCs are on average larger than at the present fit the data best. A fraction of their clusters are Roche-lobe overflowing at birth. Because their result is based on $\chi^2$ fits against the entire GC population, excluding accreted clusters, their fit results are mainly driven by clusters in the inner parts of the Milky Way. Also in our scenario D most clusters at $\rg\lesssim5\,$kpc have filled their Roche-lobes and it is hard to distinguish scenarios there. Analyses of the orbits and radii of GCs suggests that a large fraction of them are Roche-lobe filling at the present day  \citep[e.g.,][]{2010MNRAS.401.1832B, EJ2012}, supporting our conclusion that most clusters started more compact. \citet{SKYK2013} find that the CIMF must have been flat, in agreement with what we find. The flat CIMF needed to get the shape of the RD right is in stark contrast with the observed power-law mass function of young clusters \citep{2010ARA&A..48..431P} and is yet another argument that dynamical evolution cannot be responsible for the difference \citep[see also ][]{2001MNRAS.322..247V,GB2008}.


For the models described in this study we assumed the \emph{in situ} formation of the GC population. However, some of the outer halo clusters, many of which are the large outliers in the RD were most likely from dwarf satellites. A connection between dwarf galaxies and extended clusters has previously been suggested \citep{2008ApJ...672.1006E, 2009AJ....137.4361D, 2010ApJ...717L..11M}. We have furthermore made the assumption of a constant mass-loss rate, and we omitted several physical effects such as the delayed escape of stars from the Roche-lobe \citep{FH2000} and stellar evolution. A fast cluster evolution code is being developed \citep{AG2012} with the goal to include all these effects to be able to model the RD, GCMF and the $\rg$ distribution in the future.

\section*{Acknowledgements}
We thank Holger Baumgardt and Henny Lamers for insightful discussions and helpful comments. We also thank our anonymous reviewer for helpful suggestions. The authors are grateful to the Royal Society for international exchange scheme funding, and to the University of Queensland, Brisbane, where part of this study was performed. PA acknowledges financial support from STFC, and MG thanks the Royal Society for financial support.

\bibliography{radii}
 
\appendix

\end{document}